*Systems Biology*

# Enzyme promiscuity prediction using hierarchy-informed multi-label classification

Gian Marco Visani[1], Michael C. Hughes[1] and Soha Hassoun[1,2,*]

[1]Department of Computer Science, Tufts University, 161 College Ave, Medford, MA, 02155, USA,
[2]Department of Chemical and Biological Engineering, Tufts University, 4 Colby St, Medford, MA, 02155, USA

*To whom correspondence should be addressed.

**Abstract**
**Motivation:** As experimental efforts are costly and time consuming, computational characterization of enzyme capabilities is an attractive alternative. We present and evaluate several machine-learning models to predict which of 983 distinct enzymes, as defined via the Enzyme Commission (EC) numbers, are likely to interact with a given query molecule. Our data consists of enzyme-substrate interactions from the BRENDA database. Some interactions are attributed to natural selection and involve the enzyme's natural substrates. The majority of the interactions however involve non-natural substrates, thus reflecting promiscuous enzymatic activities.
**Results:** We frame this "enzyme promiscuity prediction" problem as a multi-label classification task. We maximally utilize inhibitor and unlabelled data to train prediction models that can take advantage of known hierarchical relationships between enzyme classes. We report that a hierarchical multi-label neural network, EPP-HMCNF, is the best model for solving this problem, outperforming *k*-nearest neighbours similarity-based and other machine learning models. We show that inhibitor information during training consistently improves predictive power, particularly for EPP-HMCNF. We also show that all promiscuity prediction models perform worse under a realistic data split when compared to a random data split, and when evaluating performance on non-natural substrates compared to natural substrates.

**Availability and implementation:** We provide Python code for EPP-HMCNF and other models in a repository termed EPP (Enzyme Promiscuity Prediction) at https://github.com/hassounlab/EPP.
**Contact:** soha@cs.tufts.edu
**Supplementary information:** Supplementary data are available at *Bioinformatics* online.

## 1 Introduction

Characterizing activities of enzymes plays a critical role in advancing biological and biomedical applications. While enzymes are traditionally assumed *specific*, acting on a particular molecule, many enzymes, if not all, have *promiscuous* activities acting on "non-natural" substrates, ones other than those that the enzyme evolved to transform (D'Ari and Casadesus, 1998; Khersonsky, et al., 2006; Khersonsky and Tawfik, 2010; Nobeli, et al., 2009). Despite efforts in understanding types of promiscuity (substrate vs catalytic promiscuity) and cataloguing enzyme activities on various substrates in databases, comprehensive characterization of enzyme promiscuity remains elusive.

Several applications drive the development of computational tools to analyze enzyme promiscuity. The prediction of promiscuous products of Cytochromes P450 is enabled by techniques such as Metaprint2D-react (Adams, 2010), and PROXIMAL (Yousofshahi, et al., 2015). Predicting products of metabolism, e.g., the MINEs database (Jeffryes, et al.) and BioTransformer (Djoumbou-Feunang, et al., 2019), facilitates suggesting chemical identities for compounds collected through untargeted metabolomics. The approach proposed by PROXIMAL (Yousofshahi, et al., 2015) was extended to enzymes other than Cytochromes P450 enzymes to facilitate annotation of untargeted metabolomics (Hassanpour, et al., 2020) and for suggesting enzymatic activities responsible for catalogued products in metabolomics databases (Amin, et al., 2019). Prediction of putative enzymatic links, e.g., Selenzyme (Carbonell, et al.,

2018), XTMS (Carbonell, et al., 2014), and ELP (Jiang, et al., 2020), allows the construction of novel biosynthesis or biodegradation pathways. The study of enzyme promiscuity also elucidates the evolution of metabolic networks (Carbonell, et al., 2011). Further, enzyme promiscuity can be tuned and exploited for novel protein engineering applications (Chen and Arnold, 2020).

We investigate in this paper data-driven approaches to predict enzyme classes, as defined via the Enzyme Commission (EC) numbers, that are likely to interact with a given query molecule. One approach is based on *k*-nearest neighbor (*k*-NN) similarity, where fingerprint similarity is calculated between the query molecule and the "natural" (native) substrates that are known to be catalyzed by the enzyme under natural selection and under cellular physiological conditions. The remaining four approaches are novel adaptations of several machine-learning models that frame the problem as multi-label classification, where each predicted label corresponds to an EC number. Each model has an architecture that embeds a different information-sharing paradigm across enzymes. Some models exploit hierarchical relationships encoded in the EC Nomenclature, which provides a tree-structured hierarchical classification of enzymes.

While there are works that predict enzymatic reaction similarities (Carbonell and Faulon, 2010; Rahman, et al., 2014), or use sequences to predict protein function in terms of Gene Ontology (GO) terms (Feng, et al., 2018; Kulmanov, et al., 2017; Roy, et al., 2012) or EC numbers (Concu and Cordeiro, 2019; Dalkiran, et al., 2018; Kumar and Choudhary, 2012; Li, et al., 2018; Ryu, et al., 2019), or to predict the likelihood of a sequence catalyzing a reaction or to quantify the affinity of sequences on substrates (Mellor, et al., 2016), the problem solved herein predicts enzyme classes that act promiscuously on a query molecule. To train our models, we utilize data from the BRENDA database (Schomburg, et al., 2017), which lists natural and non-natural substrates interreacting with the catalogued enzyme classes. The listed compounds for each enzyme are predominantly non-natural substrates. As we train our models using this dataset of predominantly promiscuous substrate-enzyme interacting pairs, we refer to this problem as the "enzyme promiscuity prediction" problem. Solving the enzyme promiscuity prediction problem efficiently as proposed herein allows for the quick exploration of enzyme classes that act on molecules. These findings can be further refined by identifying specific sequences that interact with the molecules and that are compatible with the host. Our best-in-class technique, EPP-HMCNF, can therefore be intergraded with a variety of tools to explore biotransformation routes (Moura, et al., 2013), assess solubility within the host (Amin, et al., 2019), or determine the likelihood of an enzyme sequence acting on a molecule borrowing from drug-target interaction prediction techniques (Chen, et al., 2018).

A major challenge when addressing this problem is the lack of available data with sufficiently representative examples of negative cases (non-interacting pairs). The BRENDA database provides cases of positive enzyme-molecule interactions. The BRENDA database also lists inhibitor molecules, those that bind with the enzyme (similar to positive cases) but where the catalytic activity is inhibited. Inhibitors thus do not uniformly represent the negative set and cannot be utilized as such. To address the lack of negative data (non-interacting enzyme-substrate pairs), we select molecules from the BRENDA database that are not in the enzyme's positive nor inhibitor lists, referred to herein as unlabeled molecules, and treat them as negatives during training and testing. This assumption is reasonable as negative interactions far outnumber positive interactions in nature. Importantly, to make judicious use of this unlabeled data during training, we apply a probabilistic weight to each such example to reflect our confidence in the negative label, deriving the weight from the molecule's structural similarity score to the most similar positive example.

Our results show that the best model for predicting enzyme promiscuity is EPP-HMCN, a model based on Hierarchical Multi-label Classification Network (known as HMCN-F, where F indicates a feed-forward architecture) (Wehrmann, et al., 2018). EPP-HMCNF provides information sharing along the EC hierarchy and across enzymes at each level of the hierarchy. We investigate the performance of EPP-HMCNF against similarity-based methods, finding that, while similarity is a competitive baseline, EPP-HMCNF delivers better performance and stands to offer further improvement as more labeled data becomes available. The main contributions of this work are:

- We develop and evaluate machine-learning classifiers with a range of different patterns of information sharing across enzymes
- We demonstrate the effective utilization of a large dataset of promiscuous interactions. We provide predictions for 983 enzyme classes, a huge increase over previous efforts that made predictions for four specific enzymes (Pertusi, et al., 2017) using SVMs and active learning.
- While substrate similarity is widely used for determining promiscuity in metabolomic engineering practice (Pertusi, et al., 2015), there is currently no large-scale systematic evaluation of the effectiveness of similarity in predicting enzyme promiscuity across enzymes or enzyme classes. Further, prior works have shown no consensus on a similarity level that deems a query molecule sufficiently similar to a native substrate. Our work offers such a study. Importantly, our work shows that machine learning outperforms similarity-based methods on several metrics including R-Precision (R-PREC), the metric that is most significant from a user's perspective as it best correlates with downstream usefulness for an experimentalist.
- Our experiments show that inhibitors are hard negative examples and that including inhibitor information during training consistently improves predictive power, particularly for EPP-HMCNF.
- Our results show that all promiscuity prediction models perform worse under a realistic data split (Martin, et al.), when compared to a random data split, and when evaluating performance on non-natural substrates compared to natural substrates.

## 2 Methods

### 2.1 Dataset

All positive and inhibitor molecules were collected from BRENDA, excluding co-factors because these metabolites are common among many enzymatic reactions. The Morgan fingerprint, with a radius of 2 and 2048 binary features (Rogers and Hahn, 2010), is used to represent each molecule. Compound names in BRENDA that could not be mapped to a specific molecular structure were discarded. By the end of the conversion process from names in BRENDA to Morgan fingerprints, we identified 25,872 positive pairings between molecules and EC numbers, based on 8,295 unique molecules. Within this same set of molecules, we also identified 13,087 inhibiting interactions, based on 2,165 unique inhibitors. Some enzymes had limited positive data. We focused on enzymes that had a minimum of 10 positive examples, so there would be enough data for both training and assessment. Using these inclusion criteria, our dataset consisted of 983 distinct EC numbers, of which 730 had at least one known inhibitor. All 983 EC numbers came from top-level classes 1-6, as we excluded the recently established top-level class 7 due to insufficient data. Of all positive pairings, only ~13% are associated with an EC class paired with a natural substrate. These pairings are associated with in vivo metabolism as specified under each EC Number's "Natural Substrates Products" section in the BRENDA database. The ratio of natural substrate to total interacting positive molecules, listed under the "Substrates/Products" section, varied per EC Number (Fig. 1).

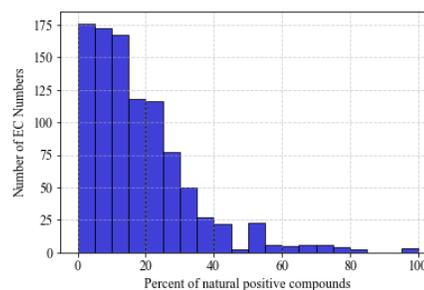

**Fig. 1.** Distribution of per EC number percentage of natural compounds to all compounds in our dataset.

For every EC number, each of the 8,295 molecules that were not known to be positive nor inhibitors are considered unlabeled in the BRENDA database. Per EC number, these unlabeled molecules were treated as negative examples during training, as were the inhibitors. Across enzymes, the ratio of positive-to-all molecules was on average 0.0032, with a standard deviation of 0.0044. The $10^{th}$ percentile of this distribution was 0.0013, while the $90^{th}$ was 0.0057.

Since inhibitors bind with enzymes, they might be considered 'closer' to the positive molecules and thus be more difficult (harder) to classify. In Machine Learning applications, including such 'hard negative' examples can fine-tune the decision boundary between positives and negatives (e.g., (Radenović, et al., 2016)). We validated that inhibitor molecules across enzymes are, on average, more similar (based on Tanimoto score) to their respective positive molecules than a similar-size randomly selected set of unlabeled molecules (Supplementary File 1, Section 1, Figure S1), thus confirming their status as hard negative examples.

Data was further organized in a tree hierarchy to match the structure of the EC nomenclature. There were 6 nodes at the class level (top of hierarchy), 50 nodes at sub-class level, 146 nodes at sub-subclass level, and 983 leaf nodes (distinct EC numbers). At all non-leaf nodes in the hierarchy, positive examples consisted of the union of all positive examples at any child. Similarly, inhibitor (hard negative) data consisted of inhibitors at any child unless already labelled positive due to a positive label from any other child node. At each node, any molecule that is not positive nor an inhibitor is considered unlabeled. In the cases in which a molecule's label is conflicted between positive and inhibitor, which may occur at internal nodes of the hierarchy, only the positive label is assigned.

## 2.2 Models

We implemented one model based on fingerprint similarity and four machine-learning models. The machine learning models (Fig. S2) are presented in the order corresponding to the amount of information sharing they allow across enzymes: no sharing (predictions for each leaf node are developed independently); top-down hierarchical sharing (each leaf uses learned representations from parents); horizontal sharing (each leaf uses learned representations common across all enzymes); and horizontal-plus-hierarchical sharing (each leaf predictor uses shared representations, as well as representations common to leaves that share a parent).

### 2.2.1 k-NN Similarity (No Sharing)

Each EC Number-molecule pair in the test set was scored by computing the mean similarity between the test molecule and the $k$ most similar positive molecules for the relevant EC Number in the training set. Similarity is evaluated via the Tanimoto score (Bajusz, et al., 2015) on the Morgan fingerprints. Hyperparameter $k$ was individually optimized for each EC Number via 3-fold cross-validation grid search on the training set, selecting the value with the best average precision score. The search considered the following values of $k$: 1, 3, 5, 7, 9, 11, 13, 15, 17, 19, 21, and 'all', where 'all' indicates that the mean similarity was computed between the test molecule and all positive molecules in the training set. A value of $k=1$ corresponds to computing the maximum similarity, while the "all" value corresponds to computing the mean similarity.

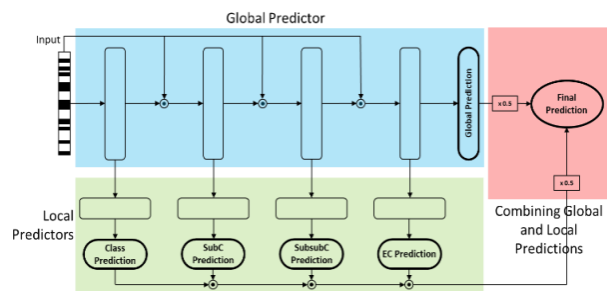

**Fig. 3. EPP-HMCNF Structure.** Global predictors are for each node in the hierarchy. Local predictors are for each level of the hierarchy. Junctions (circles with a dot inside) denote concatenation.

### 2.2.2 No-Share RF (No Sharing)

An independent Random Forest (RF) binary classifier (Breiman, 2001) was built for every leaf in the tree (distinct EC number). Each RF classifier was composed of 50 decision trees. At each internal decision node, each decision tree was trained to optimize the Gini metric while allowing for the selection of a random subset of features of size proportional to the square root of the number of features. During training, we selected model complexity hyperparameters (the minimum size of any tree's terminal node) by performing a 5-fold cross-validation grid search. The grid search considered minimum terminal node sizes of 1, 3, 5, 10, 20, 50, and 100, selecting the value with best average precision score.

### 2.2.3 Hierarchical RF (Greedy Top-Down Hierarchical Sharing)

This model (Fig. 2) uses a tree-like architecture that mimics the EC nomenclature hierarchy. A hierarchical cascade of random forests (RFs) was trained, with one RF predictor at each internal node and leaf node of the tree. Each of the 6 top-level enzyme categories had a root predictor trained to produce probabilistic predictions given data and binary labels. Then, an RF regressor at each lower-level node was trained to predict the residual error of the estimator at the parent node. The overall probabilistic prediction at a node is thus formed by adding its prediction to those from all preceding levels (thresholding to keep a valid probability in the unit interval). The hyperparameters of the RF predictors and RF regressors were selected in the same way as in No-Share RF.

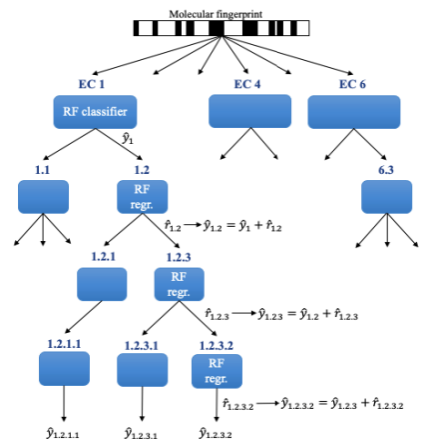

**Fig. 2. Structure of the Hierarchical RF.** RF classifiers are trained at the top-level nodes, while RF regressors are trained at the lower-level nodes. Example calculations of probabilistic predictions are shown at some nodes.

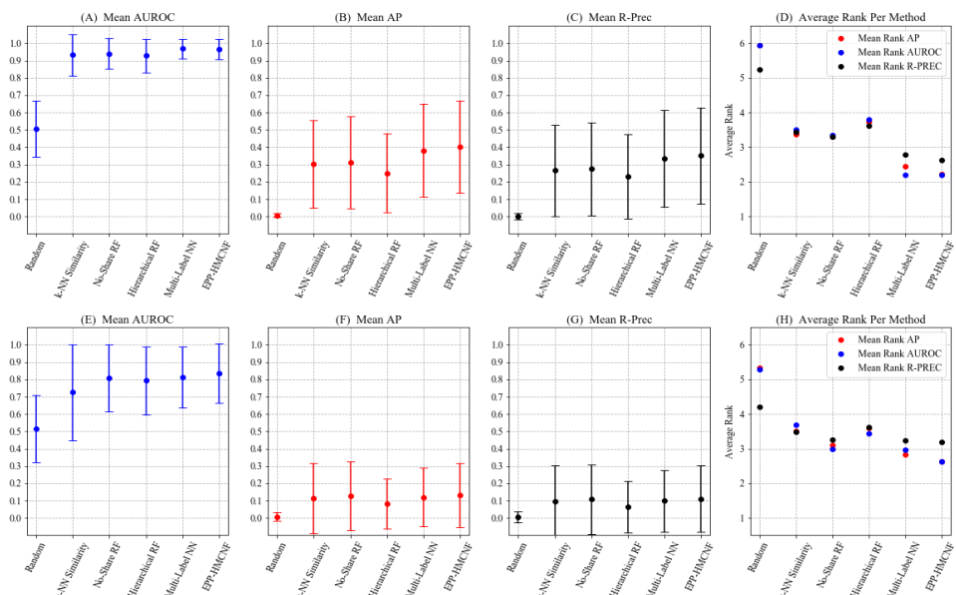

**Fig. 4. Evaluation of promiscuity prediction models trained without including inhibitor information under random split (top row), and under realistic split (bottom row), and evaluated using the Full Test Set.** (A, E) Mean AUROC, (B, F) Mean AP, (C, G) Mean R-PREC, and (D, H) Average rank per method. Intervals shown in (A, B, C, E, F, G) represent +/- 1 standard deviation.

#### 2.2.4 Multi-Label NN (Horizontal Sharing)

This model consists of a fully connected, Multi-Label Neural Network (NN) (Tsoumakas and Katakis, 2007). To make promiscuity predictions, each molecule's input features are fed through a NN with 4 hidden layers to obtain a learned representation common to all enzymes, and then a final output layer with sigmoid activation produces enzyme-specific probabilistic predictions given this common representation. In this way, all enzymes share a common feature transformation trained to improve performance across all enzymes. However, unlike Hierarchical RF, there is no specific usage of hierarchical information that would prioritize sharing between enzymes of the same class, sub-class or sub-sub-class.

Each Layer is fully connected with ReLu activation followed by batch normalization and dropout. Residual connections are present between the hidden layers. Binary cross entropy was used as loss. We used the Adam optimizer with a batch size of 12. The learning rate was manually set to 0.0002, and the model converged after 70 epochs. We cross-validated our model using random search (Bergstra and Bengio, 2012) across 12 pairs of dropout and layer size values that were sampled from uniform distributions between 0.01 and 0.6, and between 128 and 384, respectively. We selected values that maximized the average precision score.

#### 2.2.5 HMCN-F (Horizontal Plus Hierarchical Sharing)

We adapt HMCN-F, a state-of-the-art NN architecture for hierarchical multi-label classification (Wehrmann, et al., 2018) for our task. HMCN-F consists of one global predictor, which predicts all labels in the hierarchy (producing a probability for each internal node and leaf node), and local predictors for each level in the hierarchy, which predicts labels only for that specific level. The global and local predictions are then merged into a final prediction. In our adaptation (Fig. 3), our model EPP-HMCNF was implemented with 4 global hidden layers and 4 local hidden layers, one dedicated to each enzyme class. Each layer is fully connected with ReLU activation followed by batch normalization and dropout. Probabilistic predictions are made via fully connected output layers with sigmoid activation. Residual connections are present between the hidden layers in the global predictor. HMCN-F was trained similarly to the Multi-Label NN, for 70 epochs. The final prediction is obtained by equally weighting the predictions from the global and local predictors. Information is therefore shared across enzymes both in the horizontal fashion of the Multi-Label NN as well as in the vertical fashion across the EC hierarchy, as in Hierarchical RF. Unlike Hierarchical RF, however, training is not greedy across the hierarchy, but happens simultaneously at all levels.

### 2.3 Confidence Weighting of Unlabeled Data

Providing a per-example weight (a scalar positive value) to make some examples more important during training is a common technique to overcome label balance issues or to account for unlabeled data that may unknowingly contain positive examples (Liu, et al., 2003). We assign an overall weight $w_{m,e}$ to each molecule-enzyme pair $(m, e)$ that is the product of a scalar similarity weight and a scalar label weight:

$$w_{m,e} = SimW_{m,e} * LabelW_{m,e}$$

The similarity weight $SimW_{m,e}$ is used to denote our confidence in the provided positive or negative label. This value will be set to 1 for positive examples and inhibitors (since we are confident in their negative label). However, when a molecule is unlabeled for an enzyme in BRENDA, we assign it a negative label but a similarity weight between 0 and 1. This value is set to one minus the maximum similarity between the unlabeled molecule and all molecules associated with the corresponding enzyme in the positive set. Similarity is scored using the Tanimoto score for two molecular fingerprints.

To overcome the challenge of negative labels far outnumbering positive examples, all examples are given a label weight $LabelW_{m,e}$, which is the same for all examples of the same label. The label weight is set to 1.0 for positive molecule-enzyme interactions. For negative examples, it is set as:

$$LabelW_{m,e} = \frac{NumPositive_{m,e}}{\sum_m SimW_{m,e}}$$

This weight enforces that the aggregate weight of samples in the negative class is equal to that of the positive class for each enzyme, thus balancing the influence of each binary class during training.

We provide these weights when training RF predictors using the 'sample_weight' keyword argument in Scikit-Learn (Pedregosa, et al., 2011). For the NNs, we scaled the contribution of each molecule-enzyme to the loss function with its relative weight. If $\hat{y}_{m,e}$ is the predicted probability and $y$ the true binary label, then:

$$loss = \sum_{m,e} w_{m,e} * binaryCrossEntropy(y_{m,e}, \hat{y}_{m,e})$$

While we use weighted datasets for training to obtain more robust models, during model evaluation all examples are unweighted.

## 3 Results

We first trained and evaluated our models under the classic random train-test split. Using confidence weighting of unlabeled data, we trained the No-Share RF, Hierarchical RF, Multi-Label NN and HMCN-F. We then evaluated the models, together with *k*-NN Similarity. We compared our methods against a baseline 'Random' model, where every enzyme-molecule pair in the test set is assigned a likelihood of interaction value that is selected from a uniform distribution between 0 and 1.

We perform our experiments on several test sets. The "Full Test Set" comprises all enzyme-molecule paired interactions that were saved for testing, corresponding to 20% of the total interactions collected from BRENDA. The "Inhibitor Test Set" includes only the positive and inhibitor interactions, and none of the unlabeled ones, for each enzyme. The positive and inhibitor interactions were selected from the Full Test Set. Only 885 out of the 983 EC Numbers presented inhibitors, and only 671 EC Numbers presented inhibitor interactions among the data saved for testing. Therefore, only these 671 were considered for testing. The "Unlabeled Test Set" contains positive and unlabeled examples, but excludes known inhibitors, in the same positive-to-total ratio as the Inhibitor Test Set. For all test sets, unlabeled molecules are uniformly sampled within each EC Number's set. To generate a more competitive and realistic testing scenario, we split the molecular data into training and testing under a realistic split (Martin, et al., 2017), and replicated some experiments under this split.

We measured the performance of our models via three metrics: Mean Area Under the Receiver Operating Characteristic (AUROC), Mean Average Precision (AP) and Mean R-Precision (R-PREC) (Manning, 2009). Each overall summary score is respectively computed by averaging the per-EC-Number scores across all EC Numbers in the corresponding dataset. These metrics were selected as they do not require thresholding and instead consider the ranking of positive examples relative to the unknowns in the test data. This is particularly well suited for our purposes, since the similarity models do not return results that can be interpreted as probabilities, and thus thresholding would have required ad hoc and likely unjustifiable assumptions. To compute R-PREC, we are given a list of candidate molecules as input with knowledge that some number R of them is truly relevant, and then we compute the precision among the R top-scoring candidates as ranked by the predictor. A model with high R-PREC has a high probability of ranking positive molecules ahead of unknown molecules. The top-ranked molecules are naturally the first ones that the user will consider for experimental testing. Therefore, a model with high R-PREC on a test set is most likely to be useful for end users.

### 3.1 Comparing the performance of models trained without known inhibitors

To train without including inhibitor information, inhibitors are treated as unlabeled molecules and are assigned similarity-based confidence weights. Fig. 4 (A-C) show a comparison of all models for each metric. Each score is reported with +/- 1 standard deviation across all EC Numbers. The results show that *k*-NN Similarity is a strong baseline for our classification task, thus confirming that substrate similarity is a valuable tool to predict promiscuity. However, similarity is outperformed by most Machine Learning techniques. Increasing information-sharing across EC Numbers improves performance, with EPP-HMCNF showing the best mean scores across AP and R-PREC and tying with Multi-Label NN in terms of AUROC. Hierarchical RF performs worse than No-Share RF with regards to all metrics. The greedy training of Hierarchical RF may limit its ability to fix any mistakes made at higher levels of the hierarchy. Notably, the +/- 1 standard deviation intervals are overlapping across all models, indicating that the difference in performance across models is smaller than the difference across ECs.

We also measure the relative performance of our models by ranking them on each of the 983 binary classification tasks (one per EC Number) and then summarize the results by computing the average rank. Whenever two or more models tie in rank on a particular EC Number, we assign to all disputing models the average of the disputed ranks; this occurs more frequently on R-PREC scores, since its computation considers only a handful of predictions per EC Number. We repeat this for each of the three performance metrics (Fig. 4D). EPP-HMCNF ranks best for AP and R-

|  |  | AUROC | | | | AP | | | | R-PREC | | | |
|---|---|---|---|---|---|---|---|---|---|---|---|---|---|
|  |  | No-Share RF | Hierarchical RF | Multi-Label NN | EPP-HMCNF | No-Share RF | Hierarchical RF | Multi-Label NN | EPP-HMCNF | No-Share RF | Hierarchical RF | Multi-Label NN | EPP-HMCNF |
|  |  | **(A) Random Data Split** | | | | | | | | | | | |
| Mean score across ECs | No inhibitors | 0.94 | 0.928 | 0.969 | 0.965 | 0.312 | 0.251 | 0.381 | 0.402 | 0.274 | 0.23 | 0.334 | 0.351 |
|  | With inhibitors | 0.937 | 0.931 | 0.967 | 0.973 | 0.31 | 0.264 | 0.384 | 0.414 | 0.267 | 0.241 | 0.335 | 0.359 |
|  | % Difference | -0.32 | **0.32** | -0.21 | **0.83** | -0.64 | **5.18** | **0.79** | **2.99** | -2.55 | **4.78** | **0.3** | **2.28** |
| % ECs for which technique is better | No inhibitors | 49 | 45 | 52 | 41 | 50 | 45 | 49 | 43 | 51 | 48 | 50 | 47 |
|  | With inhibitors | **51** | **55** | 48 | **59** | 50 | **55** | **51** | **57** | 49 | **52** | 50 | **53** |
|  |  | **(B) Random Data Split on Reduced Dataset** | | | | | | | | | | | |
| Mean score across ECs | No inhibitors | 0.932 | 0.921 | 0.964 | 0.960 | 0.291 | 0.233 | 0.350 | 0.367 | 0.264 | 0.220 | 0.312 | 0.331 |
|  | With inhibitors | 0.928 | 0.925 | 0.962 | 0.968 | 0.290 | 0.247 | 0.353 | 0.375 | 0.255 | 0.233 | 0.317 | 0.334 |
|  | % Difference | -0.43 | **0.43** | -0.21 | **0.83** | -0.34 | **6.01** | **0.86** | **2.18** | -3.41 | **5.91** | **1.60** | **0.91** |
| % ECs for which technique is better | No inhibitors | 49 | 44 | 52 | 42 | **51** | 42 | 49 | 45 | **51** | 47 | 49 | 48 |
|  | With inhibitors | **51** | **56** | 48 | **58** | 49 | **58** | **51** | **55** | 49 | **53** | **51** | **52** |
|  |  | **(C) Realistic Data Split on Reduced Dataset** | | | | | | | | | | | |
| Mean score across ECs | No inhibitors | 0.809 | 0.793 | 0.836 | 0.849 | 0.128 | 0.083 | 0.12 | 0.131 | 0.108 | 0.063 | 0.1 | 0.111 |
|  | With inhibitors | 0.812 | 0.811 | 0.833 | 0.857 | 0.133 | 0.091 | 0.142 | 0.149 | 0.112 | 0.072 | 0.119 | 0.126 |
|  | % Difference | **0.37** | **2.27** | -0.36 | **0.94** | **3.91** | **9.64** | **18.33** | **13.74** | **3.7** | **14.29** | **19** | **13.51** |
| % ECs for which technique is better | No inhibitors | **53** | 44 | 52 | 45 | 50 | 42 | 47 | 46 | 50 | 48 | 47 | 46 |
|  | With inhibitors | 48 | **56** | 48 | **55** | 50 | **58** | **53** | **54** | 50 | **52** | **53** | **54** |

**Table 1.** Evaluation of training without and with inhibitors. (A) Under random data split on the full test set, (B) under random split on the reduced dataset (i.e. only including ECs for which models could be trained under the realistic split), and (C) under realistic split.

PREC, tying with Multi-Label NN with regards to AUROC, consolidating itself as best model for this task. Overall, the results in Fig. 4D are consistent with those in Fig. 4A-4C: models with higher scores in Fig. 4A-C generally have lower (i.e., better) rank in Fig. 4D.

We analyzed the average performance of the *k*-NN and EPP-HMCNF models for each EC Class (Supplementary File 1, section 4, Fig. S3). We found that some enzymes are easier to characterize than others. Class 3 hydrolase enzymes and Class 5 isomerase enzymes are consistently easiest to classify. Furthermore, we found that EPP-HMCNF outperforms *k*-NN similarity for all classes.

We analyze the average performance of the models across EC Numbers as a function of the number of positive examples. We evaluate the performance of EC Numbers with the most and least number of positive examples (Supplementary File 1, section 5). We show that all models have a better than average R-PREC performance and less R-PREC variability for enzymes with more labeled positive data, indicating that prediction quality is likely to increase as more data becomes available. Hierarchical RF is the model that most benefits from higher data availability, even slightly outperforming *k*-NN. While low data availability for some enzymes is partly responsible for the high variability in scores across enzymes seen in Figure 3A-C, most of the variability is best explained by suggesting that some enzymes are much easier to characterize than others, as indicated by the still high standard deviation of scores when evaluating EC Numbers with the most and least positive examples (Figure S3).

## 3.2 Training with inhibitor information

To evaluate the impact of including known negative examples in the form of inhibitors, we retrained our models including inhibitor information, setting the similarity weight to 1 for each inhibitor. We then perform two sets of experiments. First, we test the models on the "Full Test Set" and compare our findings against the result from the no-inhibitor models. Table 1(A) summarizes the differences among the models and reports the percentage of EC Numbers for which the use of each training technique yields a better score. Hierarchical RF and HMCN-F are the models that consistently benefit the most from training with inhibitors, with EPP-HMCN experiencing a significant boost for all metrics.

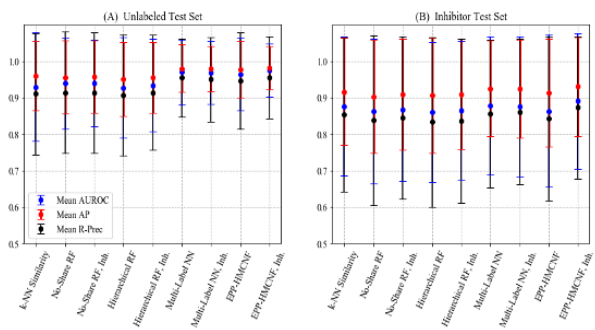

**Fig. 5. Results of training with inhibitor information.** (A) Evaluation on the Unlabeled Test set, and (B) Evaluation on the Inhibitor Test set.

Next, we considered the same metrics on the Inhibitor Test Set. The goal of using this set is to show the capacity of each model in distinguishing between positive molecule and inhibitor molecules, the only true negative examples in our dataset. We also tested our models on the Unlabeled Test Set. Since this set contains positive and unlabeled interactions with the same ratio as the ratio between positive and inhibitor interactions of the Inhibitor Test Set, any difference in scores between the Inhibitor and the Unlabeled Test Sets can be attributed entirely to group differences between the inhibitor molecules and unlabeled molecules. Results on the two test sets for all models, trained with and without inhibitor information, are shown in Fig. 5. In the Inhibitor Test Set and in the Unlabeled Test Set, the positive-to-total ratios across EC Numbers are much higher than in the Full Test Set. This is important to note since AP and R-PREC scores correlate with the positive-to-total ratio. For each of these sets, the average ratio is 0.579, with a standard deviation of 0.199, a 10[th] percentile of 0.286, and a 90[th] percentile of 0.833.

All scores are lower on the Inhibitor Test Set than on the Unlabeled Test Set, indicating that it is harder for the models to distinguish between positives and inhibitors than it is to distinguish between positives and unlabeled. The higher standard deviation also shows that the task of discerning inhibitors from positives is harder for some EC Numbers than for others. This is further confirmation that the inhibitors can be designated as "hard" negative examples. However, EPP-HMCNF trained with inhibitors is the best model at discerning inhibitors from positives (Fig. 5B), confirming EPP-HMCNF as our model of choice.

## 3.3 Realistic Split

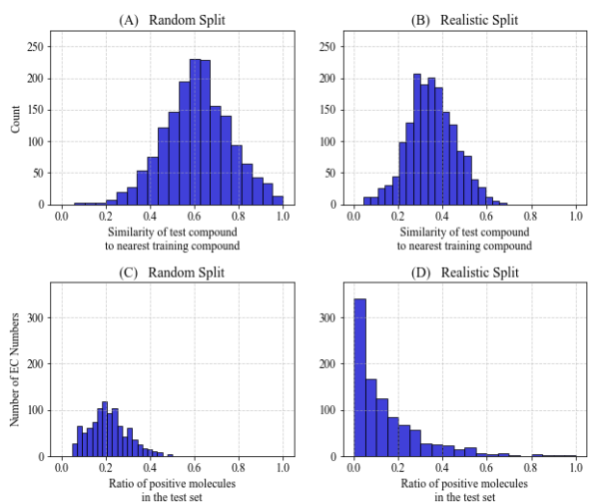

**Fig. 6. Data distributions under random vs realistic split.** (A) Histogram of maximum similarity between one test compound and all training compounds under random split. (B) Maximum similarity between one test compound and all training compounds under realistic split. (C) Distribution of *r*, the ratio of number of positive examples in test set over the number of positive examples in the training set per EC Number, under random split. (D) Distribution of *r* under realistic split.

We analyzed our models based on a "realistic data split" that contains potentially novel query compounds in comparison to the training dataset (Martin, et al., 2017). We clustered the molecules in our dataset on the molecular fingerprints with an average cluster size of 10. We used UPGMA (the unweighted pair group method with arithmetic mean), a simple bottom-up hierarchical clustering method (Sokal, et al., 1958). We saved molecules from singletons and smaller clusters, up to 20% of the total molecules, for testing. This split reduced the similarity between test and training molecules when compared to the similarity obtained under a random split (Fig. 6A-B).

The realistic split affects the distribution of positive data in the training and testing sets. Fig. 6C-D plots the distribution of the ratio $r = (\#positives\_in\_test)/(\#positivies\_in\_train)$ per EC Number under each splitting strategy. Under the random split, $r$ appears to have a unimodal, symmetric distribution centered at 20%. Under the realistic split, $r$ follows an asymmetric distribution concentrated near 0% and decaying at larger values, similar to an exponential distribution. This distribution's shape results in high number of EC Numbers with either no positive test data, or no or little training data.

Under the realistic split, models could be trained only for 680 EC Numbers, with an EC hierarchy composed of 6 Classes, 45 Subclasses, and 122 Sub-subclasses. The relative performance of the models under the random split on this reduced dataset (Table 1B) is consistent with the full test set (Table 1A) across all metrics and in terms of average ranking per method. We compare the performance on this reduced data set under the two different data splits (Table 1B and 1C). The relative performance of all models is lower under the realistic split than under the random split, with models trained with inhibitor data demonstrating slightly less degradation in mean AP and mean R-PREC. These results show that it is more difficult to obtain good model performance under the more challenging realistic split. Importantly, mean scores for almost all metrics are improved when training with inhibitors, with Multi-Label NN also experiencing a consistent boost. Crucially, EPP-HMCNF trained with inhibitor information is still the best model for the task (Mean R-PREC 0.126 vs. 0.096 of *k*-NN Similarity).

### 3.4 Predicting on Natural vs non-Natural Query molecules

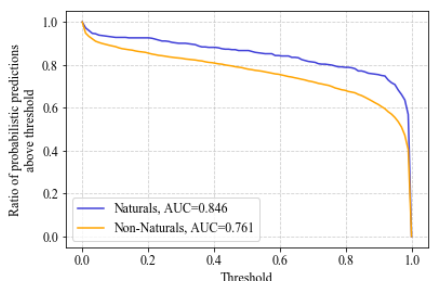

**Fig. 7. Sensitivity across thresholds on natural and non-natural test interactions.**

We evaluated the difference in performance of our best performing classifier, EPP-HMCNF trained with inhibitor information, on enzyme classes acting on natural and non-natural substrates in the Full Test Set (under the random split). For both sets of predictions, we plotted (Fig. 7) the ratio of probabilistic predictions above a series of thresholds between 0 and 1. As all test instances in this experiment are positive, higher probabilistic predictions are desirable, which translate to a higher area under the curve (AUC). The predictions on the interactions with the natural substrates have higher AUC than the predictions on the interactions with the non-natural ones (0.846 vs 0.761). We restricted the comparison to EC numbers that had both natural and non-natural test compounds. For each EC Number, we computed the mean of the predictions on the natural vs. non-natural test compounds. The AUC for the former is higher than the AUC of the latter (0.852 vs 0.761). These results suggest that it is easier for EPP-HMCNF to discern enzyme class interactions with natural substrates than with non-natural ones. Importantly, per Fig. 7, the model still classifies over 61% of non-natural compounds as positive with a probability above 90%, and over 78% of non-natural as positive with probability above 50%. This illustrates the potential of machine learning models to identify new non-natural compounds, and thus promiscuous interactions, for our EC Numbers.

### 3.5 Speed and scalability of training and inference

We report training and inference speed of all models, each with their respective optimized set of hyperparameters, and trained for all 70 epochs where applicable (Table 2). The time complexity (discussed in Supplementary File 1, Section 6) is also reported. All measurements are made when training the models on the Training set and Full Test set generated via random split. Models were trained and evaluated on an

| | Train. Time | Infer. Time | Time complexity on training | Time complexity on inference |
|---|---|---|---|---|
| k-NN Sim. | 00:00:00 | 00:12:54 | No Training | $O(ec_i \overline{p_{ec}} m)$ |
| No-Share RF | 00:12:36 | 00:00:21 | $O(ec_t \, n \log(n))$ | $O(ec_i \, m \, D)$ |
| Hierarchical RF | 00:20:48 | 00:01:22 | $O(ec_n \, n \log(n))$ | $O(ec_i \, m \, D)$ |
| Multi-Label NN | 04:06:10 | 00:00:03 | $O(n \, ec_t)$ | $O(m \, ec_t)$ |
| EPP-HMCNF | 09:38:40 | 00:00:06 | $O(n \, ec_n)$ | $O(m \, ec_n)$ |

**Table 2. Training and inference times reported in hh:mm:ss, and time complexity on training and inference, where**

$n =$ number of training molecules
$m =$ number of inference molecules
$D =$ maximum tree depth across all random forests
$p_{ec} =$ number of training positive molecules for a specific EC Number
$\overline{p_{ec}} =$ average number of training positive molecules across EC Numbers
$ec_t =$ number of EC Numbers on which the model is trained
$ec_n =$ number of nodes in the EC hierarchy used in training (note: $ec_t < ec_n \leq 4ec_t$)
$ec_i =$ number of EC Numbers for which inference is performed (note: $ec_i \leq ec_t$)

Intel(R) Xeon(R) CPU X5675 with a speed of 3.07GHz and 64GB RAM. Further speedups can be obtained via training on GPUs. *K*-NN similarity is expensive during inference, whereas, for all our dataset of 1,659 molecules, all machine learning-based models ran under 1.5 minutes, generating predictions for 983 EC Numbers. The fastest models at inference time are the NN-based models. Their runtime is independent of the number of EC Numbers of interest as the output layer calculates all predictions simultaneously, thus making them appealing for large and comprehensive sets of predictions. As more data becomes available, Multi-Label NN and EPP-HMCNF hold promise to be the fastest and most scalable inference models.

## 4  Conclusion and Discussion

This work proposed and evaluated several machine learning models to predict enzyme promiscuity on a query molecule. Our results show that sharing information both horizontally across enzymes and vertically across the EC hierarchy results in large and consistent gains in prediction quality. EPP-HMNCF implementation trained with inhibitor information is the best model for the enzyme promiscuity prediction task, achieving a Mean R-PREC of 0.359 across 983 EC numbers, against the 0.265 Mean R-PREC of *k*-NN Similarity. For a typical enzyme, slightly more than 1 in 3 wet-lab trials would succeed if we selected the top-ranked molecules using EPP-HMNCF, while a similarity-based method would yield success for 1 in 4 molecules. Thus, although similarity is a competitive baseline, it is outperformed by machine learning methods. Further, similarity is less consistent than EPP-HMCNF across EC Classes. Our data analysis and experiments show that inhibitors are hard negative examples. Indeed, EPP-HMCNF trained with inhibitor information performs with a Mean R-PREC of 0.955 on the Unlabeled Test set, while yielding a Mean R-PREC of 0.873 on the Inhibitor Test Set, indicating that inhibitors are harder to distinguish from positives when compared to unlabeled molecules. Furthermore, EPP-HMCNF trained with inhibitor information is the best model at discerning known positives from known inhibitors.

The realistic data split generates a more challenging test set than when using the traditional random split, allowing more "realistic" assessment of how well these models are likely to generalize to compounds which are different from those in the training set. EPP-HMCNF trained with inhibitor information remains the best model even on this harder task, with a mean R-PREC across 680 EC Numbers of 0.126. Thus, we expect a 1 in

8 wet lab hit rate, in contrast with the 1 in 10 hit rate provided by similarity. While it is important to evaluate models under circumstances that most closely mimic practical scenarios, splitting the data in a non-uniform way inevitably leads to per-label biasing under a multi-label setting, which may bias the models in making more accurate predictions for some labels over others. Future work should further address this issue.

Overall, our results suggest that predicting enzyme promiscuity through machine learning techniques that leverage existing knowledge in databases hold promise to advance biological engineering practices. The work presented here can be improved by integrating alternative methods from the PU learning literature (Bekker and Davis, 2018; Zhang and Zuo, 2008) and by considering learned representations that better capture molecular structure than binary fingerprint vectors, e.g. (Jin, et al., 2018).

## Acknowledgements

## Funding

This research is supported by NSF, Award CCF-1909536, and also by NIGMS of the National Institutes of Health, Award R01GM132391. The content is solely the responsibility of the authors and does not necessarily represent the official views of the National Institutes of Health.

*Conflict of Interest:* none declared.

## References

Adams, S.E.: University of Cambridge; 2010. Molecular similarity and xenobiotic metabolism.

Amin, S.A., *et al.* Towards creating an extended metabolic model (EMM) for E. coli using enzyme promiscuity prediction and metabolomics data. *Microbial Cell Factories* 2019;18(1):109.

Amin, S.A., *et al.* Establishing synthesis pathway-host compatibility via enzyme solubility. *Biotechnol Bioeng* 2019;116(6):1405-1416.

Bajusz, D., Rácz, A. and Héberger, K. Why is Tanimoto index an appropriate choice for fingerprint-based similarity calculations? *Journal of cheminformatics* 2015;7(1):20.

Bekker, J. and Davis, J. Learning from positive and unlabeled data: A survey. *arXiv preprint arXiv:1811.04820* 2018.

Bergstra, J. and Bengio, Y. Random search for hyper-parameter optimization. *Journal of machine learning research* 2012;13(Feb):281-305.

Breiman, L. Random forests. *Machine learning* 2001;45(1):5-32.

Carbonell, P. and Faulon, J.L. Molecular signatures-based prediction of enzyme promiscuity. *Bioinformatics* 2010;26(16):2012-2019.

Carbonell, P., Lecointre, G. and Faulon, J.L. Origins of specificity and promiscuity in metabolic networks. *J Biol Chem* 2011;286(51):43994-44004.

Carbonell, P., *et al.* XTMS: pathway design in an eXTended metabolic space. *Nucleic Acids Res* 2014;42(Web Server issue):W389-394.

Carbonell, P., *et al.* Selenzyme: enzyme selection tool for pathway design. *Bioinformatics* 2018;34(12):2153-2154.

Chen, K. and Arnold, F.H. Engineering new catalytic activities in enzymes. *Nature Catalysis* 2020;3(3):203-213.

Chen, R., *et al.* Machine Learning for Drug-Target Interaction Prediction. *Molecules* 2018;23(9).

Concu, R. and Cordeiro, M. Alignment-Free Method to Predict Enzyme Classes and Subclasses. *Int J Mol Sci* 2019;20(21).

D'Ari, R. and Casadesus, J. Underground metabolism. *Bioessays* 1998;20(2):181-186.

Dalkiran, A., *et al.* ECPred: a tool for the prediction of the enzymatic functions of protein sequences based on the EC nomenclature. *BMC Bioinformatics* 2018;19(1):334.

Djoumbou-Feunang, Y., *et al.* BioTransformer: a comprehensive computational tool for small molecule metabolism prediction and metabolite identification. *Journal of cheminformatics* 2019;11(1):1-25.

Feng, S., Fu, P. and Zheng, W. A hierarchical multi-label classification method based on neural networks for gene function prediction. *Biotechnology & Biotechnological Equipment* 2018;32(6):1613-1621.

Hassanpour, N., *et al.* Biological Filtering and Substrate Promiscuity Prediction for Annotating Untargeted Metabolomics. *Metabolites* 2020;10(4).

Jeffryes, J.G., *et al.* MINEs: open access databases of computationally predicted enzyme promiscuity products for untargeted metabolomics. *Journal of cheminformatics* 2015;7(1):44.

Jiang, J., Liu, L.P. and Hassoun, S. Learning graph representations of biochemical networks and its application to enzymatic link prediction. *Bioinformatics* 2020.

Jin, W., Barzilay, R. and Jaakkola, T. Junction tree variational autoencoder for molecular graph generation. *arXiv preprint arXiv:1802.04364* 2018.

Khersonsky, O., Roodveldt, C. and Tawfik, D.S. Enzyme promiscuity: evolutionary and mechanistic aspects. In.; 2006.

Khersonsky, O. and Tawfik, D.S. Enzyme promiscuity: a mechanistic and evolutionary perspective. *Annual review of biochemistry* 2010;79:471-505.

Kulmanov, M., Khan, M.A. and Hoehndorf, R. DeepGO: predicting protein functions from sequence and interactions using a deep ontology-aware classifier. *Bioinformatics* 2017;34(4):660-668.

Kumar, C. and Choudhary, A. A top-down approach to classify enzyme functional classes and sub-classes using random forest. *EURASIP J Bioinform Syst Biol* 2012;2012(1):1.

Li, Y., *et al.* DEEPre: sequence-based enzyme EC number prediction by deep learning. *Bioinformatics* 2018;34(5):760-769.

Liu, B., *et al.* Building Text Classifiers Using Positive and Unlabeled Examples. In, *ICDM*. Citeseer; 2003. p. 179-188.

Manning, C.D., Prabhakar Raghavan, and Hinrich Schutze. Evaluation in information retrieval. In, *Introduction to information retrieval*. 2009.

Martin, E.J., *et al.* Profile-QSAR 2.0: Kinase Virtual Screening Accuracy Comparable to Four-Concentration IC50s for Realistically Novel Compounds. *J Chem Inf Model* 2017;57(8):2077-2088.

Mellor, J., *et al.* Semisupervised Gaussian Process for Automated Enzyme Search. *ACS Synth Biol* 2016;5(6):518-528.

Moura, M., Broadbelt, L. and Tyo, K. Computational tools for guided discovery and engineering of metabolic pathways. In, *Systems metabolic engineering*. Springer; 2013. p. 123-147.

Nobeli, I., Favia, A.D. and Thornton, J.M. Protein promiscuity and its implications for biotechnology. *Nature biotechnology* 2009;27(2):157-167.

Pedregosa, F., *et al.* Scikit-learn: Machine learning in Python. *Journal of machine learning research* 2011;12(Oct):2825-2830.

Pertusi, D.A., *et al.* Predicting novel substrates for enzymes with minimal experimental effort with active learning. *Metab Eng* 2017;44:171-181.

Pertusi, D.A., *et al.* Efficient searching and annotation of metabolic networks using chemical similarity. *Bioinformatics* 2015;31(7):1016-1024.

Radenović, F., Tolias, G. and Chum, O. CNN image retrieval learns from BoW: Unsupervised fine-tuning with hard examples. In, *European conference on computer vision*. Springer; 2016. p. 3-20.

Rahman, S.A., *et al.* EC-BLAST: a tool to automatically search and compare enzyme reactions. *Nat Methods* 2014;11(2):171-174.

Rogers, D. and Hahn, M. Extended-connectivity fingerprints. *Journal of chemical information and modeling* 2010;50(5):742-754.

Roy, A., Yang, J. and Zhang, Y. COFACTOR: an accurate comparative algorithm for structure-based protein function annotation. *Nucleic Acids Res* 2012;40(Web Server issue):W471-477.

Ryu, J.Y., Kim, H.U. and Lee, S.Y. Deep learning enables high-quality and high-throughput prediction of enzyme commission numbers. *Proc Natl Acad Sci U S A* 2019;116(28):13996-14001.

Schomburg, I., *et al.* The BRENDA enzyme information system–From a database to an expert system. *Journal of biotechnology* 2017;261:194-206.

Sokal, R.R., Michener, C.D. and University of Kansas. A statistical method for evaluating systematic relationships. University of Kansas; 1958.

Tsoumakas, G. and Katakis, I. Multi-label classification: An overview. *International Journal of Data Warehousing and Mining (IJDWM)* 2007;3(3):1-13.

Wehrmann, J., Cerri, R. and Barros, R. Hierarchical multi-label classification networks. In, *International Conference on Machine Learning*. 2018. p. 5075-5084.

Yousofshahi, M., *et al.* PROXIMAL: a method for Prediction of Xenobiotic Metabolism. *BMC systems biology* 2015;9(1):94.

Zhang, B. and Zuo, W. Learning from positive and unlabeled examples: A survey. In, *2008 International Symposiums on Information Processing*. IEEE; 2008. p. 650-654.